\documentclass[%
 aip,
 amsmath,amssymb,
 reprint,%
]{revtex4-1}

\usepackage{graphicx}% Include figure files
\usepackage{dcolumn}% Align table columns on decimal point
\usepackage{bm}% bold math
\usepackage{booktabs}
\usepackage{multirow}
\usepackage{amsmath}
\usepackage{geometry}
\usepackage[utf8]{inputenc}
\usepackage[T1]{fontenc}
\usepackage{mathptmx}
\usepackage{etoolbox}

\usepackage{placeins}
\usepackage[authormarkuptext=name,authormarkup=none]{changes}
\definechangesauthor[name={Gianluca}, color={blue}]{GL}
\definechangesauthor[name={Thanja}, color={purple}]{TL}
\definechangesauthor[name={Hannes}, color={green}]{HJ}

\makeatletter
\def\@email#1#2{%
 \endgroup
 \patchcmd{\titleblock@produce}
  {\frontmatter@RRAPformat}
  {\frontmatter@RRAPformat{\produce@RRAP{*#1\href{mailto:#2}{#2}}}\frontmatter@RRAPformat}
  {}{}
}%
\makeatother

\begin{document}

\preprint{AIP/123-QED}

\title{
Valence and Rydberg excited state bond dissociation curves of CO$_2$ from orbital-optimized density functional calculations
}

\author{Darío Barreiro-Lage}
\affiliation{Leiden Institute of Chemistry, Gorlaeus Laboratories, Leiden University, PO Box 9502, 2300 RA Leiden, The Netherlands}
\email{d.barreiro.lage@lic.leidenuniv.nl}
\author{Gianluca Levi}
\affiliation{
Department of Chemical and Pharmaceutical Sciences, University of Trieste, 34127 Trieste, Italy}
\affiliation{Science Institute and Faculty of Physical Sciences, University of Iceland, 107 Reykjavík, Iceland}
\email{gianluca.levi@units.it}
\author{Hannes Jónsson}
\affiliation{Science Institute and Faculty of Physical Sciences, University of Iceland, 107 Reykjavík, Iceland}
\author{Thanja Lamberts}
\affiliation{Leiden Institute of Chemistry, Gorlaeus Laboratories, Leiden University, PO Box 9502, 2300 RA Leiden, The Netherlands}
\affiliation{Leiden Observatory, Leiden University, PO Box 9513, 2300 RA Leiden, The Netherlands}
\email{a.l.m.lamberts@lic.leidenuniv.nl}

\date{\today}

\begin{abstract}
Calculations of the lowest valence $\pi^*$ as well as the 3$s$ and higher energy $3p\sigma$ Rydberg excited states of the CO$_2$ molecule are carried out using density functionals with 
variational optimization of the orbitals, an approach involving relatively little computational effort.
Five functionals with varying degree of exchange are used in combination with real or complex-valued orbitals that are optimized by finding saddle points on the electronic energy surface corresponding to the excited states. 
When the PBE functional is used in combination with complex orbitals, the calculated excitation energy is found to be within 0.3 eV of multireference configuration interaction reference values, and the results are further improved with hybrid functionals.  
In contrast, linear-response time-dependent density functional theory calculations give errors up to 1.9 eV for the most diffuse $3p\sigma$ excitation and exhibit stronger dependence on both the excitation character and the functional used.  
Calculated C-O dissociation curves using the PBE functional and the orbital-optimized approach compare remarkably well with the reported multireference configuration interaction and equation-of-motion coupled-cluster singles and doubles calculations.
Thanks to the low computational cost, these results demonstrate that orbital-optimized density functional calculations can be a promising route for modelling photorelaxation in condensed-phase CO$_2$, for example in the context of interstellar cosmic-ray radiation-driven process involving high-energy Rydberg states.
\end{abstract}

 \maketitle

\section{Introduction}\label{sec:Intro}
Advances in laser technology and spectroscopy techniques have transformed the study of photochemistry, providing direct observation of excited state dynamics with near-atomic resolution.\cite{centurion2022ultrafast} 
The interpretation of the experimental observations and identification of the underlying mechanism of photoinduced processes remains highly challenging and relies in part on theoretical calculations. 
The accuracy of excited state dynamics simulations critically depends on the quality of the underlying electronic structure method \cite{Janos2023}, as illustrated by a recent prediction challenge regarding the dynamics of the cyclobutanone molecule following photoexcitation to the $3s$ Rydberg state.\cite{vindel2024predict, janovs2024predicting, mukherjee2024prediction} 
Quantum chemistry methods with high accuracy  involve large computational effort that scales rapidly with the number of electrons and are, therefore, limited to small systems.

As a computationally affordable alternative, the time-dependent extension of density functional theory (TD-DFT)\cite{Runge1984, HohenbergPhysRev.136.B864} 
is most widely used for describing excited states of molecules. However, most practical TD-DFT calculations rely on the adiabatic approximation and linear response\cite{Casida1995}, and thereby do not describe well excitations involving large rearrangement of the electron density, such as core and charge transfer excitations, as well as excitations to Rydberg states. 
When based on local and semi-local Kohn-Sham (KS)\cite{KohnShamPhysRev.140.A1133} 
functionals, TD-DFT calculations typically underestimate the excitation energy of such transitions.\cite{Maitra2017, Seidu2015, Cheng2008} 
Range-separated hybrid functionals\cite{KronikSteinRefaelyAbramsonBaer2012, Stein2009,yanai2004new} 
can alleviate this issue by incorporating Fock exchange at long range, but their accuracy depends sensitively on system- and state-specific tuning of the range separation parameters.\cite{Maitra2017} 
Nonadiabatic approximations to TD-DFT remain at an early stage of development.\cite{Lacombe2023}

Alternatively, excited states can be computed efficiently within conceptually simpler time-independent density functional approaches where the orbitals are variationally optimized for the excited state of interest.\cite{Vandaele2022, Hait2021, LeviJonsson2020_JCTC, Kowalczyk2011}
There, a solution, possibly an approximate one because of some some additional constraints introduced\cite{Pham2025, Kussmann2024, Gavnholt2008}, of the KS equations is obtained for nonaufbau occupation of the orbitals. When optimizing the orbitals of a single Slater determinant, these approaches are also sometimes referred collectively as delta self-consistent field ($\Delta$SCF), which highlights that the excitation energy is obtained as the difference between the total energy of separate SCF calculations. 

One strategy to obtain KS solutions with nonaufbau orbital occupation is to make use of eigendecomposition of the KS Hamiltonian matrix targeting solutions higher in energy than the ground state, i.e., as an extension of standard ground state SCF procedures. In practice, however, when no constraints are used, this approach often suffers from convergence problems and may fail to preserve the desired nonaufbau occupation during the SCF cycles, even when a strategy such as the maximum overlap method\cite{Gilbert2008, Barca2018} is used\cite{Schmerwitz2026, Bogo2024, Ivanov2021, CarterFenk2020, Hait2020sgm}. A more robust approach, which gives unconstrained variational optimization of the orbitals, is based on the realization that such higher order solutions to the KS equations correspond to saddle points on the electronic energy surface.
\cite{Schmerwitz2023, Hait2021} 
An excited state can thus be found by an extension of methods that were originally developed for finding first order saddle points
on potential energy surfaces.\cite{Schmerwitz2026, Schmerwitz2023, Ivanov2021, LeviJonsson2020_JCTC, LeviFaraday2020} 

Several recent studies show that state-specific orbital optimization provides a more accurate description of molecular electronic excitations that are challenging for standard TD-TDFT aproaches\cite{Vigneshwaran2025}, including core\cite{Hait2020core, Hait2020radicals}, charge transfer\cite{Schmerwitz2026, Bogo2025, Froitzheim2024, Bogo2024, Selenius2024, Barca2018, briggs2015density} and Rydberg excited states.\cite{Sigurdarson2023, Seidu2015} The performance of OO calculations using generalized gradient approximation (GGA) and meta-GGA functionals of Rydberg excited states of small molecules has been presented recently.\cite{Sigurdarson2023} Remarkably, even the commonly used PBE\cite{perdew1996generalized} functional, which is of GGA form, is found to provide good results. While it systematically underestimates the excitation energy with respect to experimental estimates, the mean absolute error is only $\sim$0.2 eV. The more elaborate meta-GGA functionals do not show significant improvement but the inclusion of explicit self-interaction correction\cite{Perdew81}
further improves the agreement with experimental estimates, likely due to the fact that it corrects the long-range form of the effective potential of the electrons. While previous benchmarks focused on the vertical excitation energy, the performance with respect to excited state potential energy surfaces of Rydberg states has not, to our knowledge, been systematically investigated. Therefore, it remains unclear how accurately the approach describes the variation of the energy of Rydberg states as the structure of the molecule changes, especially where a crossings with nearby valence excited states occur.

The CO$_2$ molecule is large enough to present a significant challenge for excited state calculations using quantum chemistry approaches. It exhibits a rich manifold of excited singlet states, some of which have mixed valence and Rydberg character.\cite{spielfiedel1992bent, england1981theoretical, knowles1988assignment} In the vacuum-ultraviolet region near the ground state equilibrium geometry, Rydberg states cross with closely lying valence states, leading to a highly corrugated electronic landscape.\cite{Lu2015, Triana2022, zhang2022photodissociation, grebenshchikov2012crossing, Grebenshchikov2013} The excited states of CO$_2$ are, therefore, an ideal test case.  
Triana et al.\cite{Triana2022} as well as Lu et al.\cite{Lu2015} have reported excited state energy curves obtained by employing the complete active space self-consistent field (CASSCF) method followed by a multireference configuration interaction (MRCI) calculation. The calculations were found to undergo spurious oscillations when the active space does not include the Rydberg orbitals. In contrast, equation of motion coupled-cluster singles and doubles (EOM-CCSD) calculations provide smooth energy curves in the Franck–Condon region, but do not have the correct dissociation limits. To obtain high-level reference curves, Triana et al.\cite{Triana2022} therefore employ a hybrid strategy, smoothly matching an EOM-CCSD energy curve in the Franck–Condon region to an MRCI energy curve at larger distance between the atoms.

It is often assumed that excitation to the UV-accessible CO$_2$ states leads to fast internal conversion followed by dissociation into CO + O.\cite{Triana2022, Grebenshchikov2013} 
However, recent studies show that some of the higher-energy Rydberg states are not dissociative and can trap the excited state population over a timescale of $\sim$150 fs.\cite{Triana2022} While such a short timescale may seem irrelevant for chemical processes, it may be sufficient to strongly affect neighboring molecules, especially given the highly diffuse nature of the Rydberg states. Its transient nature and potential role in delaying dissociation make it particularly relevant in astrochemical environments, where
condensed phase CO$_2$ has been detected in space as interstellar and planetary ices\cite{brunken2024jwst}. Furthermore, radiation-driven processes and transient intermediates play a key role in molecular evolution. Indeed, cosmic radiation leads to the formation of free electrons ($<20$ eV) in the ice, which can excite CO$_2$ molecules into these high-energy Rydberg states. The effect of, \emph{e.g.}, repulsion between the Rydberg excited molecule and its neighbors is currently not taken into consideration in astrochemical models\cite{Donoghue2022}. 

In the calculations presented here, the focus is on the 3$s$ and $3p\sigma$ Rydberg states and the lowest $\pi^*$ valence excited state of $\rm CO_2$. The 3$s$ Rydberg state is dissociative and of particular interest because it is the energetically lowest Rydberg excitation and, close to the ground state equilibrium geometry, exhibits crossings with the $\pi^*$ state. By contrast, the $3p\sigma$ state lies higher in energy and remains bound over the full range of internuclear distances and has therefore been proposed to be responsible for trapping of excited state population after UV excitation. Owing to their diffuse character and diverse potential energy surfaces, these states provide prototypical cases for benchmarking computational approaches targeting both Rydberg and valence excitations.

The article is organized as follows. Section II describes the methodology, illustrates the importance of using complex orbitals, and presents an assessment of the basis set. The first part of section III presents the results of the calculations of the vertical excitation energy of the valence and Rydberg states of CO$_2$ using various functionals as well as comparison to higher-level EOM-CCSD results. In the second part of section III, the calculated excited state dissociation energy curves along the C–O coordinate are analyzed and compared to previously reported EOM-CCSD/MRCI calculations. Discussions of the results are presented in sections IV. Section V summarizes the main conclusions.

\section{Methodology}
\subsection{Direct Orbital Optimization}
The calculations are carried out using
direct optimization (DO), where the orbitals are optimized by finding the unitary transformation of an initial set of orbitals that makes the KS energy stationary.\cite{Ivanov2021, LeviJonsson2020_JCTC, LeviFaraday2020, Voorhis2002, Head-Gordon1988} The initial set is generated from some reference set of orthonormal molecular orbitals, typically the ground state orbitals with occupations changed to reflect the desired excitation. 
The unitary transformation is parameterized by the exponential of an anti-Hermitian rotation matrix. The saddle point search can be carried out either by using efficient approximate second-order quasi-Newton methods\cite{Schmerwitz2026, Ivanov2021, LeviJonsson2020_JCTC} or by inverting the components of the gradient along the eigenvectors 
of the $n$ lowest eigenvalues when an $n$-order saddle point is the targeted.\cite{Schmerwitz2023}
Here, convergence on the saddle points is obtained with the L-SR1 quasi-Newton algorithm.\cite{Ivanov2021, LeviJonsson2020_JCTC}

\begin{figure*}[!ht]
    \centering
    \includegraphics[width=1\linewidth]{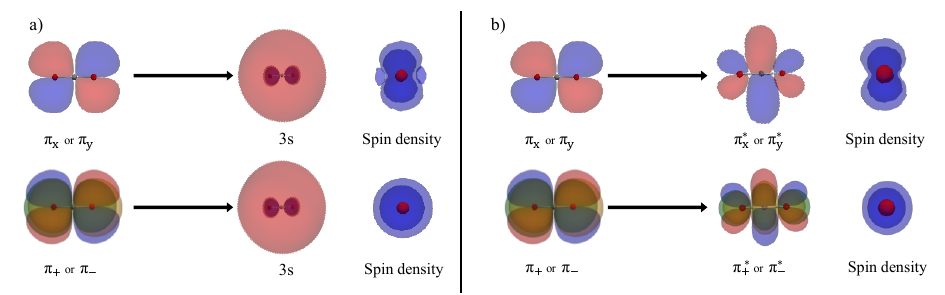 }
    \caption{
    Illustration of the advantage of using complex orbitals when calculating excitations from the occupied $\pi$ orbitals of CO$_2$ to the $3s$ orbital, shown in (a), and to the $\pi^{*}$ orbital, shown in (b).
    The $\pi$ and $\pi^{*}$ orbitals (side view) as well as the spin density of the 
    excited state (viewed along the molecular axis) 
    is shown for calculations using real orbitals, $\pi_x$ or $\pi_y$ (upper part), 
    and complex orbitals, $\pi_+$ or $\pi_-$ (lower part).
    When real orbitals are used, the cylindrical symmetry of the excited state is broken because they are not 
    eigenstates of the angular momentum operator.
      The orbitals are visualized at an isovalue of 0.02~a$_{0}^{-3/2}$, while the spin density is visualized at an isovalue of 0.002~a$_{0}^{-3}$.}
    \label{fig:Orbitals}
\end{figure*}

Since the calculations are based on a single determinant, the open-shell singlet states are spin mixed. The energy values are, therefore, corrected by applying spin purification\cite{Ziegler1977}
\begin{equation}\label{eq:spin_purification}
    E_{\mathrm{s}} = 2E_{\mathrm{sm}} - E_{\mathrm{t}},
\end{equation}
where \(E_{\mathrm{sm}}\) is the energy of the spin-mixed solution and \(E_{\mathrm{t}}\) is the energy of the corresponding triplet state. Consequently, two separate calculations are performed, using analogous nonaufbau occupations of the orbitals.

\subsection{Computational Settings}
The calculations are performed within the $D_{\infty h}$ symmetry group, keeping the bond angle at $\theta= 180.0$º. The vertical excitation energy is calculated for the experimentally determined structure, with a C-O bond length of 1.162 \AA. \cite{herzberg1966molecular} 

The OO calculations of the excitation energy and C-O dissociation energy curves are carried out with complex orbitals and the PBE functional using the Grid-based Projector Augmented Wave (GPAW) software.\cite{Mortensen2024} There, the PAW formalism\cite{Blochl1994} is used to describe the region close to the nuclei based on the frozen-core approximation. Valence electrons are represented with the d-aug-cc-pVDZ basis set, from which uncontracted functions are removed. A grid spacing of 0.15 \r{A} is used and the cell is chosen such that the distance between the outermost atoms and the cell boundaries in all three directions is 10 \r{A} to avoid any effect due to truncation of the numerical representation of the basis functions (see also section~\ref{sec:basisset}). 
The GPAW software can also be used for condensed phase simulations as it can incorporate periodic boundary conditions.

For comparison, calculations using linear-response TD-DFT are carried out with various hybrid functionals, including PBE0\cite{adamo1999toward}, B3LYP\cite{Lee1988, Becke1988}, BHHLYP\cite{becke1993new}, and CAM-B3LYP\cite{yanai2004new}, in addition to the PBE functional. There, real orbitals are used together with the d-aug-cc-pVDZ\cite{woon1994a} basis set taken from the Basis Set Exchange\cite{schuchardt2007a, feller1996a}. All calculations of excitation energy using real orbitals are performed with the ORCA 6.1\cite{ORCA6} software. 

For the calculations of the C-O bond dissociation energy curves, the same simulation box size is used for all points. To this end, a calculation at the largest CO + O separation is first carried out, applying a 10 \r{A} vacuum layer as described above. The same cell dimensions are then used for all points along the dissociation curve. 
Each excited state scan is carried out in two segments starting from the experimental geometry of the ground state of CO$_2$: one toward dissociation and the other toward shorter C–O distances, with a step size of 0.02 \r{A}. 
All subsequent points along the scan use the previously converged excited state solution as the initial guess. 

\subsection{Complex orbitals}\label{sec:complex}
For all the excited states considered here, the excitation is from a degenerate pair of $\pi$ orbitals. 
Some of the excitations are to a degenerate pair of $\pi^{*}$ orbitals.
Proper eigenstates of the angular momentum operator with cylindrical symmetry need to be described using complex orbitals,
$\pi_{\pm}$ for the ground states and $\pi^{*}_{\pm}$ for the excited state.
Such calculations can be carried out with the GPAW software.
The ORCA calculations, however, are necessarily limited to real orbitals, $\pi_x$ and $\pi_y$,
and do not preserve the linear point-group symmetry of the electron density\cite{Ivanov2021}, as illustrated in Fig.~\ref{fig:Orbitals}. 
The relationship between the two is
\begin{equation}
    \pi_{\pm} = \frac{1}{\sqrt{2}}\left(\pi_{\mathrm{x}} \pm i \, \pi_{\mathrm{y}}\right),
\end{equation}
and similarly for the $\pi^{*}$ orbitals, see also Appendix~\ref{sec:AppendixA}.
Fig.~\ref{fig:Orbitals} shows how the complex $\pi_{\pm}$ and $\pi^{*}_{\pm}$ orbitals preserve the cylindrical symmetry of the spin density for the $3s$ and $\pi^{*}$ excited states, while the real orbitals do not.
The GPAW calculations of the excitation energy are carried out for both real and complex orbitals using the PBE functional.

\subsection{Assessment of the basis set }\label{sec:basisset}
\begin{figure}[!t]
    \centering
    \includegraphics[width=1\linewidth]{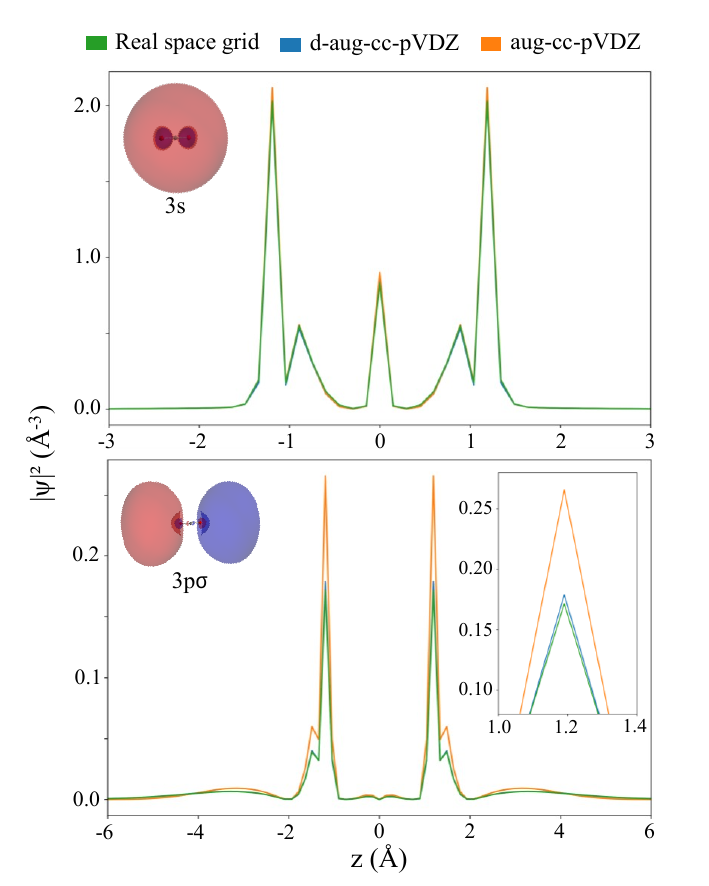}
    \caption{Comparison of results obtained with linear combination of atomic orbitals basis sets and a real space grid representation in PBE calculations of the $3s$ (top) and $3p\sigma$ (bottom) Rydberg states. The orbital densities computed with the d-aug-cc-pVDZ basis set closely match the real space grid densities. For the $3s$ state, the two orbitals are nearly indistinguishable, and for the more diffuse $3p\sigma$ state only a small deviation is observed (see inset). By contrast, the aug-cc-pVDZ orbital densities, being more confined, show clear discrepancies for the $3p\sigma$ state.}
    \label{fig:fdvslcao}
\end{figure}
It is challenging to represent Rydberg orbitals with a linear combination of atomic orbitals (LCAO) because of their highly diffuse electron distribution.\cite{Sigurdarson2023} Therefore, it is important to asses the description of the Rydberg orbitals with the employed LCAO basis set.
Figure \ref{fig:fdvslcao} compares the magnitude of the $3s$ and $3p\sigma$ orbitals obtained using a real space grid representation with results obtained with the aug-cc-pVDZ and d-aug-cc-pVDZ\cite{dunning1989a, kendall1992a} basis sets.
The latter has an extra diffuse function and thus can better represent the Rydberg states. The calculations employ the PBE functional with complex orbitals and are carried out with the GPAW software. The calculations using the aug-cc-pVDZ basis set produce orbitals that are significantly more localized compared to the real space grid representation, especially for the $3p\sigma$ state. In contrast, when the d-aug-cc-pVDZ basis set is employed, the orbitals closely reproduce the real space grid representation. Table ~\ref{table: table1} lists the values of excitation energy computed with the different basis sets. The LCAO basis sets yield systematically higher excitation energy due to confinement of the orbital. The difference in excitation energy between the real space grid representation and the aug-cc-pVDZ basis set is large for the $3p\sigma$ state, $\sim0.5$ eV. When using d-aug-cc-pVDZ, the deviation is much smaller, only 0.03 eV for the 3$s$ state and 0.1 eV for the more diffuse $3p\sigma$ state.

\begin{table}[h!]
\centering
\begin{tabular}{c|c|c}
\hline
\textbf{} & $\Delta E_{3s}$ (eV) & $\Delta E_{3p\sigma}$ (eV) \\
\hline
aug-cc-pVDZ & 8.73 & 11.44 \\
d-aug-cc-pVDZ & 8.67 & 11.11 \\
Real space grid & 8.64 & 10.98 \\
\hline
\end{tabular}
\caption{Comparison of the excitation energy (eV) of the $3s$ and $3p\sigma$ states obtained for two different atomic orbitals basis sets and a real space grid.
Clearly, the aug-cc-pVDZ basis set does not include diffuse enough functions while calculations using the d-aug-cc-pVDZ basis set reproduce better the real space grid results.}
\label{table: table1}
\end{table}

In section III, the calculated excited state energy curves of CO$_2$ are compared to the reference EOM-CCSD/MRCI curves obtained by Triana \textit{et al}\cite{Triana2022} with a basis set of atomic natural orbitals (ANO-L) augmented with diffuse \textit{s}, \textit{p} and \textit{d} functions. The excitation energy values obtained here with the d-aug-cc-pVDZ basis set are found to deviate by at most 0.01 eV from those obtained in PBE calculations with the same ANO-L basis set used in Triana \textit{et al}.\cite{Triana2022}

\section{Results}
\subsection{Excitation energy}
\begin{figure*}[!t]
    \centering
    \includegraphics[width=0.9\linewidth]{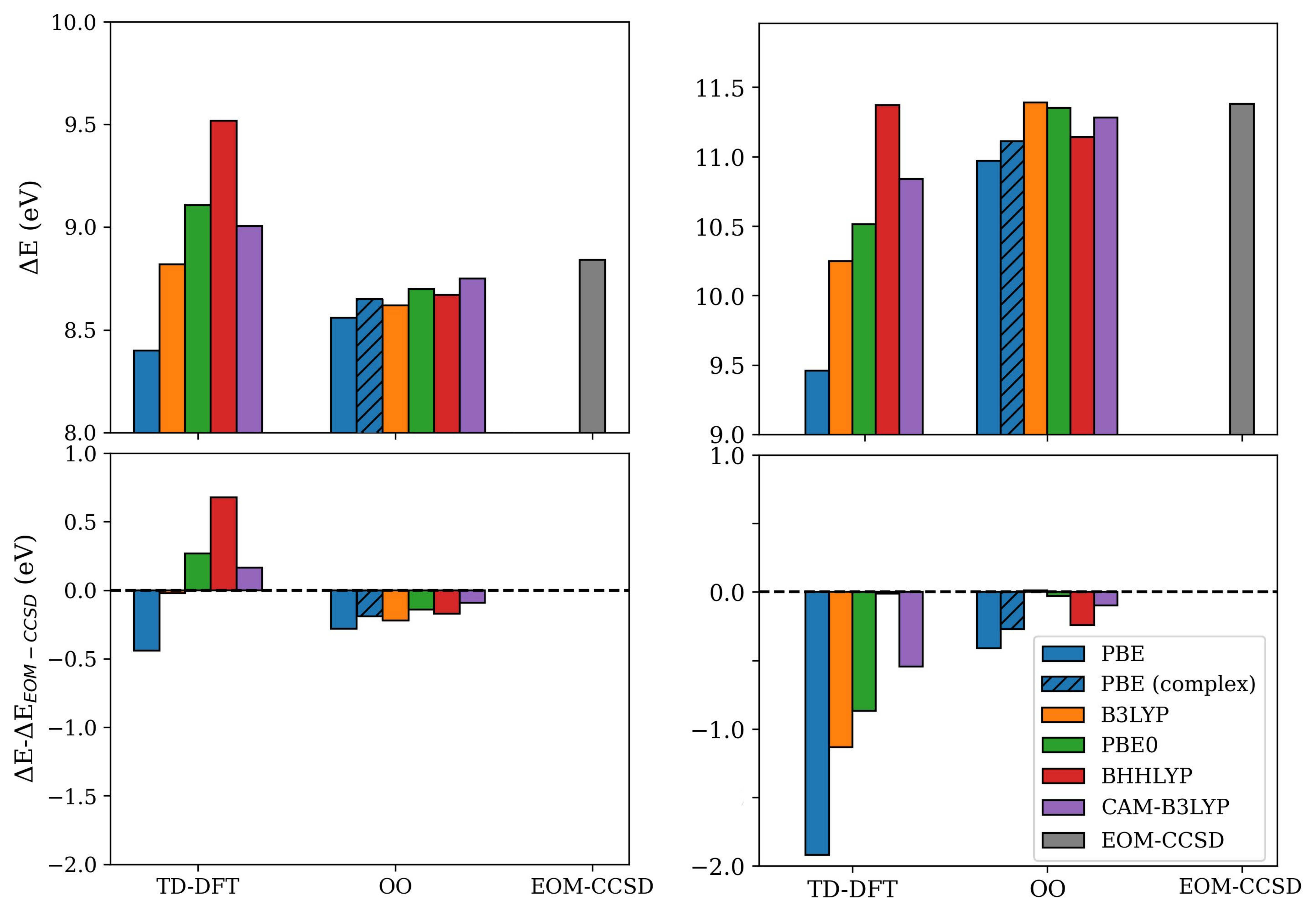}
    \caption{Excitation energy for the Rydberg excited 3$s$ (left) and $3p\sigma$ (right) states of CO$_2$ obtained with OO and TD-DFT calculations as well as reference EOM-CCSD values from Triana et al.\cite{Triana2022} The calculations presented here use the d-aug-cc-pVDZ basis set, while the EOM-CCSD calculations use an ANO-L basis set augmented with diffuse $s$, $p$, and $d$ functions localized on the carbon atom. The upper graphs illustrate the calculated excitation energy, $\Delta$E, while the lower graphs show the deviation from EOM-CCSD. The calculated values are listed in Table~\ref{app:table} in Appendix~\ref{sec:AppendixB}. The TD-DFT results exhibit a pronounced dependence on both the functional and the character of the excitation, with the largest errors observed for the more diffuse $3p\sigma$ state, whereas the OO results are less functional dependent and generally in closer agreement with the EOM-CCSD results.
}
    \label{fig:benchmark}
\end{figure*}

Figure~\ref{fig:benchmark} shows the excitation energy of the Rydberg $3s$ and $3p\sigma$ states of CO$_2$. These states have symmetry $^1\Pi_g$ and $^1\Pi_u$ and arise from a $3s \leftarrow \pi$ and $3p\sigma \leftarrow \pi$ exitation, respectively. Results obtained with OO and TD-DFT calculations using various density functionals (PBE, B3LYP, PBE0, BHHLYP, and CAM-B3LYP) are compared against the EOM-CCSD reference values reported by Triana \textit{et al.}\cite{Triana2022}

TD-DFT calculations with the PBE functional underestimate the excitation energy for both states, with an absolute error of about 0.5 eV for the $3s$ state and nearly 2 eV for the more diffuse $3p\sigma$ state. 

When global hybrid functionals are used, TD-DFT predicts higher excitation energy. This can be explained by the fact that the inclusion of Fock exchange partially corrects the self-interaction error, thereby improving the long-range form of the potential and stabilizing the occupied orbitals relative to the virtual ones. Correspondingly, the excitation energy increases with the fraction of Fock exchange. For the more compact $3s$ state, B3LYP gives the most accurate results, whereas BHHLYP, which has a larger fraction of Fock exchange, overestimates the excitation energy by more than 0.5 eV, and therefore has the largest error of the functionals tested. In contrast, the more diffuse $3p\sigma$ state requires larger fraction of exact exchange. There, BHHLYP has the smallest error, while the excitation energy remains underestimated by more than 1 eV in the B3LYP calculations. For both excitations, PBE0 gives results that are in between the B3LYP and BHHLYP results, consistent with an intermediate weight on Fock exchange. TD-DFT calculations with the range-separated hybrid functional CAM-B3LYP performs well for the $3s$ state but underestimates the excitation energy of the $3p\sigma$ state by $\sim$0.5 eV. 
Overall, the excitation energy calculated with TD-DFT spans a wide range of values, and the error depends strongly on both the functional and the diffuseness of the excited state.

For the OO calculations, the excitation energy is systematically underestimated, but the absolute deviation from the EOM-CCSD reference remains below 0.5 eV in all cases. The largest error is exhibited by the PBE functional when using real orbitals. The use of complex orbitals, which are more accurate since they have the right symmetry, provides better results. The results are further improved with the global hybrid and the range-separated CAM-B3LYP functionals. Overall, OO calculations give results in closer agreement with the EOM-CCSD reference with an error that depends less on both the functional and the excitation character than the TD-DFT calculations. 

Calculations using the PBE functional and complex orbitals offer a favorable balance between computational cost and accuracy in the OO approach. Since a future aim is to calculate excitations of CO$_2$ in condensed phase, this methodology is used in the following assessment of excited state dissociation energy curves.

\subsection{Excited state dissociation energy curves}
\begin{figure*}[!t]
    \centering
    \includegraphics[width=0.9\linewidth]{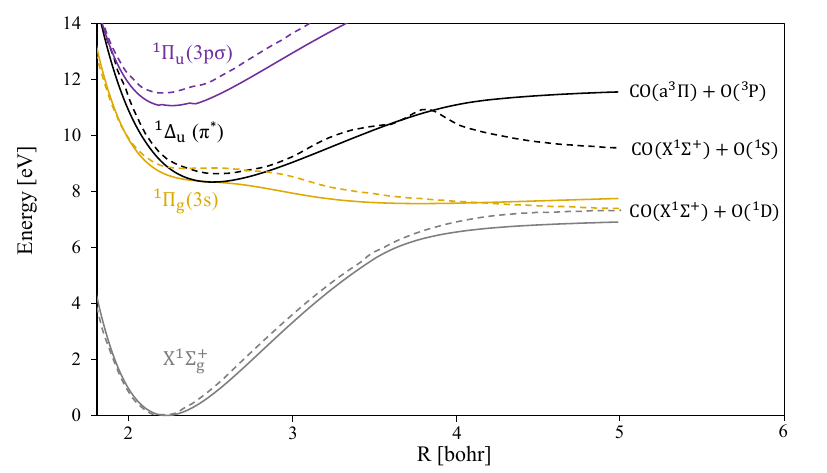}
    \caption{Energy as a function of C–O distance for one of the bonds in the linear CO$_2$ molecule in the ground state and three excited states: the 3$s$ and $3p\sigma$ Rydberg states and the lowest valence excited state, $\pi^*$. Solid lines show results obtained with the orbital-optimized (OO) approach with PBE functional and complex orbitals. 
    Dashed lines correspond to reference diabatic energy curves reported by Triana \textit{et al.}\cite{Triana2022} constructed by smoothly joining equation-of-motion coupled-cluster singles and doubles calculations in the Franck–Condon region with multireference configuration interaction results at extended bond lengths for $C_s$ symmetry (OCO angle of 179.99 deg.). 
    Symmetry labels correspond to the $D_{\infty h}$ point group, while the states in the reference calculations all belong to the A$^\prime$ irreducible representation. The OO curves correspond to diabatic representation. In the Franck–Condon region, the OO curves reproduce well the overall shape of the reference curves and are only uniformly shifted to lower energy by $\sim0.5$ eV. At larger distance, the OO $\pi^*$ curve deviates from the reference. This might be explained by the diabatization of the reference curves not including an upper state potentially involved in an avoided crossing.}
    \label{fig:scan_ryd}
\end{figure*}

Figure~\ref{fig:scan_ryd} shows the potential energy curves for a C-O bond dissociation computed with the OO approach using the PBE functional and complex orbitals. The curves for the ground state and the excited $3s$, $3p\sigma$, and $\pi^*$ states are shown. The latter arises from a $\pi^* \leftarrow \pi$ excitation and has $^1\Delta_u$ symmetry (see Appendix \ref{sec:AppendixA}). The curves are compared to reference dissociation curves reported by Triana \textit{et al.} which are obtained by smoothly joining MRCI curves in the Franck–Condon region with EOM-CCSD curves at extended bond length, and finally applying a pseudo-diabatization procedure.\cite{Triana2022}
In the OO calculations, the CO$_2$ molecule has $D_{\infty h}$ point group symmetry (OCO angle = 180 deg.). Instead, the reference diabatic curves were obtained for the $C_s$ point group (OCO angle = 179.99 deg.).\cite{Triana2022} For the $D_{\infty h}$ symmetry, the $^1\Pi_g(3s)$, $^1\Delta_u(\pi^*)$, and $^1\Pi_u(3p\sigma)$ states are doubly degenerate, whereas they split into quasi-degenerate A$^\prime$ and A$^{\prime\prime}$ states in the $C_s$ point group. In Figure~\ref{fig:scan_ryd}, the OO curves are compared to the reference EOM-CCSD/MRCI curves obtained for the A$^\prime$ states.

Close to the Franck-Condon region, the OO/PBE calculations underestimate the energy of all three excited states. However, the underestimation is approximately constant along the CO bond distance, $R$, and the magnitude of the error is similar for all excited states. Thus, the OO approach with the PBE functional provides a good description of the shape of the energy curves as well as the energy separation between the excited states around the Franck-Condon region. It appears that the OO-calculated curves approximate the curves of diabatic states, and reproduce the double crossing between the Rydberg 3$s$ and the valence $\pi^*$ states in the range of 2.4-2.7 Bohr, although the gap between the two states is underestimated around 2.5 Bohr, due to an underestimation of the barrier for the 3$s$ state. 

To verify the spin character of the fragments at dissociation, a Bader analysis of the spin density is performed,
\begin{equation}
\rho_\mathrm{s}(\mathbf{r}) = \rho_\alpha(\mathbf{r}) - \rho_\beta(\mathbf{r}),
\end{equation}
integrating $\rho_\mathrm{s}(\mathbf{r})$ over the Bader basins, $\Omega_A$, defined from the total electron density, $\rho(\mathbf r)=\rho_\alpha(\mathbf r)+\rho_\beta(\mathbf r)$:
\begin{equation}
M_A = \int_{\Omega_A} \rho_s(\mathbf{r}) \, d\mathbf{r}.
\end{equation}
where $M_A$ is a local spin population.

For the 3$s$ Rydberg excitation, both the OO and the reference calculations predict dissociation into two singlet states (CO($X\,^1\Sigma^+$) and O($^1D$)), the lowest-energy dissociation channel in CO$_2$. 
Both the ground state and the 3$s$ Rydberg state should converge to the same dissociation limit. However, the OO results deviate from this expected behavior: At 4.2 Bohr, the OO 3$s$ curve crosses the corresponding reference curve, and for larger CO bond distances, it overestimates the energy of the 3$s$ state.
The spin-mixed solution corresponding to the 3$s$ state is found to exhibit a dissociation behavior closer to the expected one. This suggests that the spin purification procedure for the $3s$ state may no longer be reliable at large CO bond separations.

The $3p\sigma$ state remains bound over the whole range in C-O distance, consistent with the EOM-CCSD/MRCI reference. The error in the excitation energy is slightly larger than for the $3s$ and $\pi^*$ states, but the shape of the curve is reproduced well. 

The dissociation energy curve of the $\pi^*$ valence state obtained with OO calculations with PBE functional departs markedly from the corresponding reference curve beyond $R=3.8$ Bohr. For the OO/PBE calculations, the energy continues to rise and only begins to level off near 5 Bohr, whereas the reference curve reaches a maximum at $R \approx 3.8$ Bohr and subsequently decreases. For the OO/PBE calculation, elongation along the CO bond in the $\pi^*$ state leads to dissociation into two fragments with triplet spin multiplicity (CO($a^3\Pi$) + O($^3$P)) as indicated by a Bader analysis of the spin density. For the reference calculations, however, the molecule dissociates to form CO($X^1\Sigma^+$) + O($^1$S). The OO/PBE curve seems to correspond to diabatic representation over the whole range in CO bond distance, as indicated by the crossings with the $3s$ state around the Franck-Condon region. The smooth maximum in the reference curve is consistent with an avoided crossing between the $\pi^*$ state and a higher-lying state of the same $A^\prime$ symmetry, in the $C_s$ point group. While the reference curves were obtained using diabatization, the upper state of the avoided crossing might not have been included explicitly. The resulting ``diabatic'' curve would then correspond to the adiabatic representation beyond the avoided crossing. This provides a possible explanation for the divergence between the OO/PBE results and the reference curves for the $\pi^*$ state as the molecule dissociates.

\begin{figure}
    \centering
     \includegraphics[width=1.1\linewidth]{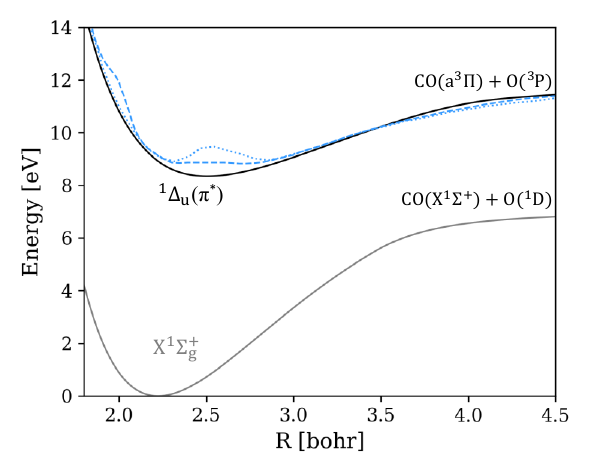}
    \caption{Energy of the $\pi^*$ excited state of CO$_2$ as a function of the C-O distance of one of the bonds
    in the CO$_2$ molecule. The results obtained with orbital-optimized (OO) calculations using the PBE functional and complex orbitals (solid line) are compared with multireference configuration interaction calculations using two different active spaces: a (16,12) active space including only valence orbitals (dotted line, results from Lu et al.\cite{Lu2015}), and a smaller (12,11) active space including the 3$s$ Rydberg orbital (dashed lines, reported by Triana et al.\cite{Triana2022}). Symmetry labels correspond to the $D_{\infty h}$ point group, while the states in the reference calculations belong to the A$^{\prime\prime}$ irreducible representation. The OO/PBE curve approximates a curve in the diabatic representation, while the MRCI curves correspond to adiabatic representation. Overall, the OO calculations reproduce remarkably well the energy curve of the $\pi^*$ state, except in the region close to 2.5 Bohr, where crossing with the 3$s$ Rydberg state occurs (see also Figure \ref{fig:scan_ryd}).
}
    \label{fig:scan_pi}
\end{figure}
Figure \ref{fig:scan_pi} compares the OO/PBE energy curve of the $\pi^*$ state (already shown in Figure \ref{fig:scan_ryd}) with adiabatic energy curves obtained by Triana et al.\cite{Triana2022} entirely from MRCI calculations for the $\pi^*$ state of A$^{\prime\prime}$ symmetry. MRCI results using two different active spaces are shown: a (16,12) active space including only valence orbitals and a (12,11) active space including the 3$s$ Rydberg orbital but fewer valence orbitals. Unlike the A$^{\prime}$ state shown in Figure \ref{fig:scan_ryd}, the A$^{\prime\prime}$ state dissociates into CO($a^3\Pi$) + O($^3$P). Overall, the OO curve agrees remarkably well with the MRCI curves except around 2.5 Bohr where the $\pi^*$ and 3$s$ states cross, as the OO curve behaves diabatically, while the MRCI curves are adiabatic. This figure also illustrates the challenges faced by multireference methods. When the Rydberg 3$s$ orbital is not included in the active space, the barrier near 2.5 Bohr is overestimated. However, when the Rydberg 3$s$ orbital is included but the active space overall reduced, a spurious bump is observed at short distances. These issues do not affect the OO calculations, as they do not rely on the definition of an active space.

\section{Discussion}
As shown above, the OO calculations with PBE describe the Rydberg 3$s$ and $3p\sigma$ as well as the valence $\pi^*$ excited states of CO$_2$ with a similar accuracy. This is a significant advantage over TD-DFT, where calculations of the different states with the same functional can yield errors that vary considerably, complicating the construction of reliable potential energy surfaces and hampering application in excited state dynamics simulations. The OO results are less sensitive to the choice of functional, with errors in all cases below 0.5 eV, thanks to the fact that the orbitals are optimized for each state separately.\cite{Vigneshwaran2025} 

The OO approach reproduces the shape of the potential energy curves of the three states remarkably well relative to the reference curves obtained with the hybrid EOM-CCSD and MRCI approach, particularly in the Franck-Condon region. Importantly, the OO calculations require only a fraction of the computational effort of EOM-CCSD and MRCI calculations. This makes applications to condensed-phase systems possible. Extending the OO approach to excited states of CO$_2$ clusters will enable studies of the atomic-scale mechanisms of light- and cosmic ray induced reactions in interstellar ices. Such studies will represent a significant advance over current time-dependent gas-grain chemical models\cite{shingledecker2018cosmic, Holdship2017}, which depend on parameters derived from experimental data\cite{shingledeckerHerbst2018, KellerRudek2013, Fueki1963}. In particular, it will be possible to treat state-specific reactivity and energetically induced desorption, effects that are largely absent from current modelling frameworks but may be important for excited state astrochemical processes. 

While the PBE functional in combination with the state-specific orbital optimization gives good results for the excited states of the CO$_2$ molecule, the tendency for overdelocalization within the generalized gradient approximation may have to be taken into account for systems with two or more CO$_2$ molecules are present. The hole can be expected to spread over two or more molecules. Such overdelocalization has been obtained in calculations of Rydberg states of molecular amine clusters.\cite{Gudmundsdottir14} The basic reason for this is the self-interaction error in KS functionals that results from the use of total electron density in the estimate of the classical Coulomb interaction between the electrons. When explicit self-interaction correction~\cite{Perdew81,Sigurdarson2023} is applied to the PBE funtional, proper localization consistent with experimental observations is obtained.\cite{Gudmundsdottir14} A similar approach is likely needed in studies of the excited states of CO$_2$ clusters and condensed phase.

Due to the single-determinantal nature of the density functional calculations, complex orbitals are needed to represent excitations between pairs of degenerate orbitals without breaking the cylindrical symmetry of the excited state spin density. In some cases, the single-determinantal representation prevents the calculation of excited states of a given symmetry even when complex orbitals are used. This limitation is illustrated by the $^1\Sigma^+_u$ and $^1\Sigma^-_u$ states arising from $\pi^* \leftarrow \pi$ excitation in linear CO$_2$ (see Appendix \ref{sec:AppendixA}). Although this needs to be taken into account for small, highly symmetric molecules, it is expected to be less restrictive in larger systems, particularly when an environment is included, because symmetry breaking typically lifts degeneracies.

Future work will be aimed at assessing orbital-optimized density functional calculations of excited states of molecules in the solid state, with relevance to photoinduced astrochemical processes. Since the orbitals are variationally optimized, analytical atomic forces can readily be obtained, and the calculations can, therefore, be extended to simulations of the dynamics of photoexcited molecules in solids.

\section{Conclusions}
State-specific orbital-optimized calculations using various density functionals are found to give remarkably accurate results for valence $\pi^*$ and Rydberg $3s$ and $3p\sigma$ excited states of CO$_2$. While the excitation energy is systematically underestimated compared to high-level EOM-CCSD results, the error is below $0.5$ eV across all tested functionals, with hybrid functionals providing better results than the PBE functional, which is of the GGA form. In contrast, TD-DFT shows a much larger dependence on the functional and the character of the excitation, with the excitation energy of the more diffuse $3p\sigma$ Rydberg state being severely underestimated by PBE as well as the hybrid B3LYP and PBE0 functionals. The smaller sensitivity of the OO approach to the functional and excitation character is a result of the state-specific orbital optimization.

The energy curves along the C-O bond dissociation coordinate are, furthermore, described well for the $\pi^*$, $3s$ and $3p\sigma$ states when complex orbitals are used in comination with the PBE functional.
The error with respect to reference curves obtained by merging EOM-CCSD curves at short distances with MRCI curves at long distances\cite{Triana2022} is approximately constant and is similar for the different states. As a result, the shape of the potential energy curves is reproduced well, even close to where the $\pi^*$ and $3s$ states cross. The fact that the OO curves are nearly uniformly shifted with respect to the higher-level EOM-CCSD/MRCI reference curves indicates that the error is systematic, and could come from an imbalance in the self-interaction error between the ground and excited states, as observed in previoous calculations of excited states of the ethylene molecule.\cite{Schmerwitz2022}

\begin{acknowledgments}
This work benefited from funding provided by COST (European Cooperation in Science and Technology) through Action CA20129, “Multiscale Irradiation and Chemistry Driven Processes and Related Technologies.”. TL and DBL acknowledges support from the Leids Kerkhoven-Bosscha Fonds (LKBF). This work was supported by the Icelandic Research Fund (grant nos.\ 2511544 and 2410644) and the University of Iceland Research Fund. GL acknowledges support from the ERC under the European Union's Horizon Europe research and innovation programme (grant no. 101166044, project NEXUS). Views and opinions expressed are however those of the author(s) only and do not necessarily reflect those of the European Union or ERC Executive Agency. Neither the European Union nor the granting authority can be held responsible for them. Computer resources, data storage, and user support were provided by the Icelandic Research e-Infrastructure (IREI), funded by the Icelandic Infrastructure Fund. We gratefully acknowledge Prof. José Luis Sanz-Vicario for providing the ANO-L basis set augmented with diffuse s, p and d functions employed in their study\cite{Triana2022}. 
\end{acknowledgments}

\section*{Author Declarations}
The authors have no conflicts to disclose.

\section*{Data Availability Statement}
The data that support the findings of this study are available from the corresponding author upon reasonable request. 

\section*{Bibliography}
\bibliographystyle{apsrev4-1}
\bibliography{aipsamp}

\appendix

\section{Energy of CO$_2$ excited states 
in $\boldsymbol{D_{\infty \mathrm{h}}}$ from single-determinant calculations}\label{sec:AppendixA}

In the following, the wave functions of the excited states of the CO$_2$ molecule are expressed using the complex orbitals
\begin{align}
\pi_{\pm} = \frac{1}{\sqrt{2}}\left(\pi_{x} \pm i\,\pi_{y}\right), 
\qquad
\pi^{*}_{\pm} = \frac{1}{\sqrt{2}}\left(\pi^{*}_{x} \pm i\,\pi^{*}_{y}\right).
\end{align}
In this basis, the projection of the orbital angular momentum on the molecular axis is $\lambda = \pm 1$ for each $\pi$ or $\pi^*$ orbital. The CO$_2$ molecule is assumed linear ($D_{\infty \mathrm{h}}$ point group), where the $\pi_{\pm}$ orbitals form a degenerate pair, and only open-shell orbitals are considered.

\subsection*{$\boldsymbol{(1\pi)^3(3s)^1}$ excited configuration}
In the Rydberg excited state configuration $(1\pi)^3(3s)^1$, the open-shell orbitals are 
the degenerate pair of $\pi_{\pm}$ orbitals and the nondegenerate $3s$ orbital. The symmetry-adapted many-body singlet and triplet wave functions of this state are given by 
\begin{align}
\Psi({}^1\Pi_g) =
\left\{
\begin{array}{l}
\displaystyle
\big(\,|\pi_{+}\,\overline{3s}\rangle
     + |3s\,\overline{\pi_{+}}\rangle\,\big)/\sqrt{2}, \\[6pt]
\displaystyle
\big(\,|\pi_{-}\,\overline{3s}\rangle
     + |3s\,\overline{\pi_{-}}\rangle\,\big)/\sqrt{2},
\end{array}
\right.
\end{align}
\begin{align}
\Psi({}^3\Pi_g) =
\left\{
\begin{array}{l}
\displaystyle
\big(\,|\pi_{+}\,\overline{3s}\rangle
     - |3s\,\overline{\pi_{+}}\rangle\,\big)/\sqrt{2}, \\[6pt]
\displaystyle
\big(\,|\pi_{-}\,\overline{3s}\rangle
     - |3s\,\overline{\pi_{-}}\rangle\,\big)/\sqrt{2}.
\end{array}
\right.
\end{align}
Neglecting orbital relaxation, the $\,|\pi_{+}\,\overline{3s}\rangle$ Slater determinant is an equal mixture of the singlet and triplet many-body wave function. Thus, 
\begin{align}
E\!\left(\big|\pi_{+}\,\overline{3s}\big\rangle\right)
=
\frac{1}{2}\Big(
E({}^1\Pi_g) + E({}^3\Pi_g)
\Big),
\end{align}
and the usual spin-purification formula\cite{Ziegler1977} can be used to obtain the energy of $\Psi({}^1\Pi_g)$:
\begin{align}
E({}^1\Pi_g)
=
2\,E\!\left(\big|\pi_{+}\,\overline{3s}\big\rangle\right)
-
E({}^3\Pi_g).
\end{align}

\subsection*{$\boldsymbol{(1\pi)^3(3p\sigma)^1}$ excited configuration}
For the $(1\pi)^3(3p\sigma)^1$ Rydberg excited configuration the situation is analogous, with the Rydberg $3p\sigma$ orbital replacing $3s$. The wave function of the singlet state is
\begin{align}
\Psi({}^1\Pi_u) =
\left\{
\begin{array}{l}
\displaystyle
\big(\,|\pi_{+}\,\overline{3p\sigma}\rangle
     + |3p\sigma\,\overline{\pi_{+}}\rangle\,\big)/\sqrt{2}, \\[6pt]
\displaystyle
\big(\,|\pi_{-}\,\overline{3p\sigma}\rangle
     + |3p\sigma\,\overline{\pi_{-}}\rangle\,\big)/\sqrt{2},
\end{array}
\right.
\end{align}
and the energy is given by the spin-purification formula:
\begin{align}
E({}^1\Pi_u)
=
2\,E\!\left(\big|\pi_{+}\,\overline{3p\sigma}\big\rangle\right)
-
E({}^3\Pi_u).
\end{align}

\subsection*{$\boldsymbol{(1\pi)^3(\pi^*)^1}$ excited configuration}
For the valence excited configuration $(1\pi)^3(1\pi^*)^1$, the open-shell orbitals are the degenerate pairs of $\pi_{\pm}$ and $\pi_{\pm}^*$ orbitals. Such configuration gives rise to $\Delta_u$, $\Sigma^+_u$, and $\Sigma^-_u$ states. \\

The wave function of the singlet $\Delta_u$ state (component of the total orbital angular momentum along the internuclear axis $\Lambda=\pm2$) is
\begin{align}
\Psi({}^1\Delta_u) =
\left\{
\begin{array}{l}
\displaystyle
\left( \big|\pi_{+}\,\overline{\pi^{*}_{+}}\big\rangle
      + \big|\pi^{*}_{+}\,\overline{\pi_{+}}\big\rangle \right)/\sqrt{2}, \\[6pt]
\displaystyle
\left( \big|\pi_{-}\,\overline{\pi^{*}_{-}}\big\rangle
      + \big|\pi^{*}_{-}\,\overline{\pi_{-}}\big\rangle \right)/\sqrt{2}.
\end{array}
\right.
\end{align}
$\big|\pi_{+}\,\overline{\pi^{*}_{+}}\big\rangle$ is again a mixture of the singlet and triplet wave functions. Therefore, also in this case, the spin-purification formula can be applied to obtain the energy of the singlet wave function:
\begin{align}
E({}^1\Delta_u) = 2\,E(\big|\pi_{+}\,\overline{\pi^{*}_{+}}\big\rangle) - E({}^3\Delta_u).
\end{align}

The symmetry–adapted many-body wave functions of the singlet and triplet $\Sigma^+_u$ and $\Sigma^-_u$ (component of the total orbital angular momentum along the internuclear axis $\Lambda=0$) can be written as:
\begin{align}
\Psi({}^1\Sigma_u^{+}) =
\frac{1}{2}\Big(
|\pi_{+}\,\overline{\pi^{*}_{-}}\rangle
+|\pi^{*}_{-}\,\overline{\pi_{+}}\rangle
+|\pi_{-}\,\overline{\pi^{*}_{+}}\rangle
+|\pi^{*}_{+}\,\overline{\pi_{-}}\rangle
\Big), \\[6pt]
\displaystyle
\Psi({}^1\Sigma_u^{-}) =
\frac{1}{2}\Big(
|\pi_{+}\,\overline{\pi^{*}_{-}}\rangle
+|\pi^{*}_{-}\,\overline{\pi_{+}}\rangle
-|\pi_{-}\,\overline{\pi^{*}_{+}}\rangle
-|\pi^{*}_{+}\,\overline{\pi_{-}}\rangle
\Big),\\[6pt]
\displaystyle
\Psi({}^3\Sigma_u^{+}) =
\frac{1}{2}\Big(
|\pi_{+}\,\overline{\pi^{*}_{-}}\rangle
-|\pi^{*}_{-}\,\overline{\pi_{+}}\rangle
+|\pi_{-}\,\overline{\pi^{*}_{+}}\rangle
-|\pi^{*}_{+}\,\overline{\pi_{-}}\rangle
\Big),\\[6pt]
\displaystyle
\Psi({}^3\Sigma_u^{-}) =
\frac{1}{2}\Big(
|\pi_{+}\,\overline{\pi^{*}_{-}}\rangle
-|\pi^{*}_{-}\,\overline{\pi_{+}}\rangle
-|\pi_{-}\,\overline{\pi^{*}_{+}}\rangle
+|\pi^{*}_{+}\,\overline{\pi_{-}}\rangle
\Big).
\end{align}
Based on these equations, the determinant $|\pi_{+}\,\overline{\pi^{*}_{-}}\rangle$ can be seen to be an equal–weight linear combination of all four $\Sigma_u$
states,
\begin{align}
\big|\pi_{+}\,\overline{\pi^{*}_{-}}\big\rangle
=
\frac{1}{2}\left(
\Psi\big({}^1\Sigma_u^{+}\big)
+
\Psi\big({}^1\Sigma_u^{-}\big)
+
\Psi\big({}^3\Sigma_u^{+}\big)
+
\Psi\big({}^3\Sigma_u^{-}\big)
\right), 
\end{align}
meaning that $|\pi_{+}\,\overline{\pi^{*}_{-}}\rangle$ breaks both spin and spatial symmetry of the wave function. Note that, owing to the cylindrical symmetry of the system and the fact that $|\pi^{*}_{+}(\mathbf r)|^{2} = |\pi^{*}_{-}(\mathbf r)|^{2}$, the determinant $|\pi_{+}\,\overline{\pi^{*}_{-}}\rangle$ is degenerate with $|\pi_{+}\,\overline{\pi^{*}_{+}}\rangle$. The energy of this broken–symmetry determinant is the average of the energy of the four wave functions:
\begin{align}
E\!\left(\big|\pi_{+}\,\overline{\pi^{*}_{-}}\big\rangle\right)
=
\frac{1}{4}\left(
E\big({}^1\Sigma_u^{+}\big)
+
E\big({}^1\Sigma_u^{-}\big)
+
E\big({}^3\Sigma_u^{+}\big)
+
E\big({}^3\Sigma_u^{-}\big)
\right).
\end{align}
Consequently, within the a single-determinant approach, one can obtain only the energy of the ${}^1\Delta_u$ state for the $(1\pi)^3(1\pi^*)^1$ configuration\cite{Ziegler1977}.

\begin{table*}[t]
\centering
\caption{Excitation energy for the $3s$ and 3$p\sigma$ states of CO$_2$ obtained with orbital-optimized and TD-DFT calculations using various density functionals. The excitation energy is given after spin purification according to eq \ref{eq:spin_purification}.}
\begin{tabular}{lllllllll}
\toprule
\multicolumn{1}{c}{State} & \multicolumn{1}{c}{Method} & Basis set & Remarks & PBE & B3LYP & PBE0 & BHHLYP & CAM-B3LYP \\
\hline
\multirow{3}{*}{$^{1}\Pi_{g}$(3$s$)} 
& TD-DFT & d-aug-cc-pVDZ & real orbitals             & 8.18  & 8.56 & 8.83 & 9.19 & 8.72 \\
& OO     & d-aug-cc-pVDZ & real orbitals    & 8.56  & 8.62 & 8.70 & 8.67 & 8.75 \\
& OO     & d-aug-cc-pVDZ & complex orbitals & 8.67  & --   & --   & --   & --   \\
\hline
\hline
\multirow{3}{*}{$^{1}\Pi_{g}$(3$p\sigma$)}
& TD-DFT & d-aug-cc-pVDZ & real orbitals             & 9.46  & 10.25 & 10.51 & 11.37 & 10.84 \\
& OO     & d-aug-cc-pVDZ & real orbitals    & 10.97 & 11.39 & 11.35 & 11.14 & 11.28 \\
& OO     & d-aug-cc-pVDZ & complex orbitals & 11.11 & --    & --    & --    & --    \\
\bottomrule
\label{app:table}
\end{tabular}
\end{table*}

\section{Calculated excitation energy}\label{sec:AppendixB}

\end{document}